\documentclass[aps, prd, amsmath, floats, floatfix, twocolumn,
superscriptaddress, nofootinbib, showpacs]{revtex4-1}

\usepackage{graphicx}
\usepackage{color}
\usepackage{soul}
\usepackage{url}
\usepackage{bm}         
\usepackage{times}
\usepackage{dcolumn}
\usepackage{bm}
\usepackage{epsf}
\usepackage{amssymb}

\newcommand{\beq}{\begin{equation}}
\newcommand{\eeq}{\end{equation}}
\newcommand{\beqn}{\begin{eqnarray}}
\newcommand{\eeqn}{\end{eqnarray}}

\usepackage{color}

\newcommand{\Caltech}{\affiliation{TAPIR, Walter Burke Institute for Theoretical Physics, MC 350-17,
    California Institute of Technology, Pasadena, California 91125, USA}}

\newcommand{\Cornell}{\affiliation{Cornell Center for Astrophysics and
    Planetary Science, Cornell University, Ithaca, New York, 14853, USA}}
\newcommand{\WSU}{\affiliation{Department of Physics \& Astronomy,
	Washington State University, Pullman, Washington 99164, USA}}
\newcommand{\CITA}{\affiliation{Canadian Institute for Theoretical 
    Astrophysics, University of Toronto, Toronto, Ontario M5S 3H8, Canada}}

\newcommand{\AEI}{\affiliation{Max Planck Institute for Gravitational Physics (Albert Einstein Institute), D-14476 Potsdam-Golm, Germany}}

\newcommand{\UNH}{\affiliation {Department of Physics, University of New Hampshire, 9 Library Way, Durham NH 03824, USA}}

\begin{document}

\title{Evaluating radiation transport errors in merger simulations using a Monte-Carlo algorithm}

\author{F. Foucart}\UNH,
\author{M.D. Duez}\WSU,
\author{L. E. Kidder}\Cornell,
\author{R. Nguyen}\UNH
\author{H.P. Pfeiffer}\AEI\CITA,
\author{M.A. Scheel}\Caltech

\begin{abstract}
Neutrino-matter interactions play an important role in the post-merger evolution of neutron star-neutron star and black hole-neutron star mergers.
Most notably, they determine the properties of the bright optical/infrared transients observable after a merger. Unfortunately, Boltzmann's equations of radiation transport remain too costly to be evolved directly in merger simulations. Simulations rely instead on approximate transport algorithms with unquantified modeling errors. In this paper, we use for the first time a time-dependent general relativistic Monte-Carlo (MC) algorithm to solve Boltzmann's equations and estimate important properties of the neutrino distribution function $\sim 10\,{\rm ms}$ after a neutron star merger. We do not fully couple the MC algorithm to the fluid evolution, but use a short evolution of the merger remnant to critically assess errors in our approximate gray two-moment transport scheme. We demonstrate that the analytical closure used by the moment scheme is highly inaccurate in the polar regions, but performs well elsewhere. While the average energy of polar neutrinos is reasonably well captured by the two-moment scheme, estimates for the neutrino energy become less accurate at lower latitudes. The two-moment formalism also overestimates the density of neutrinos in the polar regions by $\sim 50\%$, and underestimates the neutrino pair-annihilation rate at the poles by factors of $2-3$. Although the latter is significantly more accurate than one might have expected before this study, our results indicate that predictions for the properties of polar outflows and for the creation of a baryon-free region at the poles are likely to be affected by errors in the two-moment scheme, thus limiting our ability to reliably model kilonovae and gamma-ray bursts. 
\end{abstract}

\pacs{04.25.dg, 04.40.Dg, 26.30.Hj, 98.70.-f}

\maketitle

\section{Introduction}
\label{sec:intro}

The recent detection by the LIGO-Virgo Collaboration of gravitational waves (GWs) powered by a pair of merging neutron stars (GW170817~\cite{TheLIGOScientific:2017qsa}), followed by electromagnetic (EM) observations of the same system by a wide range of
ground-based and space-based telescopes~\cite{2017ApJ...848L..12A}, represents a major breakthrough for multi-messenger astronomy.
This event also shows the current limits of our ability to reliably extract information about merging compact objects using EM observations.
For example, GW170817 was followed by a bright {\it kilonova}~\cite{Cowperthwaite:2017dyu}, an optical/infrared transient powered
by radioactive decays in the neutron-rich ejecta produced by the merger~\cite{1976ApJ...210..549L,Li:1998bw,Roberts2011,Kasen:2013xka,Tanaka:2013ana}. EM observations of that kilonova have been used to infer plausible properties of the ejecta, and the outcome of r-process nucleosynthesis in the outflows~\cite{2017Natur.551...80K}. Determining the properties of the merging objects from those of the ejecta, however, remains difficult. 

To better constrain the properties of the merging objects from kilonova observations, we require numerical simulations of neutron star mergers capable of accurately predicting the properties of the ejected material. Unfortunately, despite rapid improvements in the accuracy and physical realism of these simulations, a few important issues are still limiting our ability to make such predictions. 

A first problem is that outflows are produced both during the merger ({\it dynamical ejecta}), and over the much slower evolution of the post-merger accretion disk ({\it disk outflows}), thus requiring simulations covering a wide range of time scales and length scales. Only a few simulations so far have attempted to self-consistently include both phases of the evolution~\cite{Fernandez:2014b,Just2014,Fernandez:2016sbf}. 

Another issue is that magnetic fields play a critical role in driving disk outflows, and in the post-merger evolution of the system~\cite{Kiuchi:2015qua,2016arXiv160300501L,Siegel:2017nub}. Yet these fields grow from very small scale instabilities that current simulations do not properly resolve~\cite{Kiuchi:2015qua}. So far, these effects have only been approximately captured through the use of sub-grid models~\cite{2014arXiv1410.0013G,Shibata:2017xht,2017ApJ...838L...2R}, or of unphysically large initial magnetic fields (see e.g.~\cite{BaiottiReview2016,Paschalidis2017} for reviews of the field). 

Finally, neutrino-matter interactions play a critical role in the evolution of the composition of the outflows~\cite{2008ApJ...679L.117SR,Wanajo2014}, are the main source of cooling in the post-merger remnant, can drive disk winds~\cite{Dessart2009,Just2014}, and deposit a large amount of energy in the polar regions through pair annihilation~\cite{1999ApJ...517..859S,Birkl2007,just:16,Fujibayashi:2017puw,2017JPhG...44h4007P}. As Boltzmann's equations of radiation transport remain too costly to include in 3D simulations, however, we rely on approximate transport methods to estimate the impact of neutrino-matter interactions in merger remnants.

Here, we focus on the latter issue. In particular, we note that approximate transport algorithms currently used in merger simulations come with potentially significant and, more importantly, so far unquantified errors. To work towards a more rigorous determination of error budgets in simulations, we use our recently developed general relativistic MC radiation transport code~\cite{Foucart:2017mbt} to evolve over a small time interval the remnant of a binary neutron star merger. We do not fully couple the MC evolution to the fluid evolution: neutrino-matter interactions are still handled using our approximate, gray two-moment scheme~\cite{Foucart:2015gaa,Foucart:2015vpa,Foucart:2016rxm} with M1 closure~\cite{Minerbo1978}. Instead, we use the MC evolution to obtain more accurate estimates of the neutrino distribution function. This allows us to constrain the accuracy of simulations that use approximate two-moment transport schemes. Such error estimates are critical to our ability to assess the robustness of kilonova models, as simulations using the two-moment scheme are being used to model merger outflows and to interpret kilonova observations.  

This MC simulation also provides us with an opportunity to study more carefully the properties of neutrinos emitted by the remnant of neutron star mergers. For the first time, we have at our disposal an estimate of the 7-dimensional neutrino distribution function, $f_{(\nu_i)}(t,x^i,p_i)$ (with $t$ the time, $x^i$ the spatial coordinates, and $p_i$ the spatial components of the 4-momentum of the neutrinos) for each neutrino species $\nu_i$, obtained from a time-dependent evolution of Boltzmann's equations, in general relativity and with a realistic background for the metric and fluid properties. Knowing $f_{(\nu_i)}$, rather than its lowest moments in momentum space, is important for accurate estimates of the rate of $\nu \bar\nu$ annihilations~\cite{1999ApJ...517..859S,Fujibayashi:2017puw}, and for studies of more complex neutrino processes currently not included in merger simulations (e.g. neutrino oscillations~\cite{2012PhRvD..86h5015M,2016PhRvD..93d5021M,2016PhLB..752...89W,2016PhRvD..94j5006Z}). We illustrate this advantage of the MC scheme by providing the energy spectrum of neutrinos, their angular distribution at selected points in the simulation, and the heating rate of the fluid due to $\nu \bar\nu$ annihilation in the polar regions. 

\section{Methods}

\subsection{Evolution algorithm}

In this work, we evolve the remnant of a binary neutron star merger with our general relativistic radiation hydrodynamics code, SpEC~\cite{SpECwebsite}. SpEC
evolves Einstein's equations of general relativity on a pseudo-spectral grid, using the generalized harmonic formalism~\cite{Lindblom:2007}. The general relativistic equations of hydrodynamics are evolved on a separate numerical grid using high-order shock capturing finite volume methods~\cite{Duez:2008rb}. A more detailed description of our latest methods to evolve the metric and fluid in SpEC can be found in~\cite{Foucart:2013a}.

We also evolve the general relativistic equations of neutrino radiation transport using a gray (i.e. energy-integrated) two-moment formalism~\cite{1981MNRAS.194..439T,shibata:11}.  We only use 3 distinct species of neutrinos: electron neutrinos $\nu_e$, electron antineutrinos $\bar\nu_e$, and heavy-lepton neutrinos $\nu_x$. The latter class includes muon and tau (anti)neutrinos, ($\nu_\mu,\bar\nu_\mu,\nu_\tau,\bar\nu_\tau$). At the densities and temperatures encountered in neutron star mergers, the fraction of heavy-leptons ($\mu,\tau$) in the fluid is negligible, and the 4 species gathered in $\nu_x$ are largely interchangeable. A detailed description of our implementation of the moment formalism is provided in~\cite{FoucartM1:2015,Foucart:2016rxm}. Here, we limit ourselves to a discussion of the most important aspects of the algorithm for the purpose of estimating errors in the two-moment formalism.

In the moment formalism, we evolve the lowest moments of the neutrino distribution function $f_{(\nu_i)}(t,x^i,p_i)$ in momentum space. In a coordinate system comoving with the fluid, the $0^{\rm th}$, $1^{\rm st}$, and $2^{\rm nd}$ moments are the energy density $J$, momentum density $H^\mu$ and
pressure tensor $S^{\mu\nu}$. These moments can be explicitly written as the momentum-space integrals
\beqn
J_{(\nu_i)} &=& \int d\nu \nu^3 \int d\Omega f_{(\nu_i)},\\
H_{(\nu_i)}^\mu &=& \int d\nu \nu^3 \int d\Omega f_{(\nu_i)} l^\mu,\\
S_{(\nu_i)}^{\mu\nu} &=& \int d\nu \nu^3 \int d\Omega f_{(\nu_i)} l^\mu l^\nu,
\eeqn
with $\nu$ the neutrino energy in the fluid frame, $\int d\Omega$ an integral over solid angle on a unit sphere in momentum space, and
\beq
p^\mu = \nu (u^\mu + l^\mu),
\eeq
the 4-momentum of neutrinos, with $u^\mu$ the 4-velocity of the fluid and $l^\mu u_\mu=0$. The stress-energy tensor of the neutrinos is then, for species $\nu_i$,
\beq
T^{\mu\nu}_{(\nu_i)} = J_{(\nu_i)} u^\mu u^\nu + H_{(\nu_i)}^\mu u^\nu + H_{(\nu_i)}^\nu u^\mu + S_{(\nu_i)}^{\mu\nu}.
\eeq
In simulations, we also define the energy density $E_{(\nu_i)}$, momentum density $F_{(\nu_i),\mu}$ and pressure tensor $P_{(\nu_i),\mu\nu}$ measured by an observer whose worldline is tangent to $n^\mu$, the unit normal to a $t={\rm constant}$ hypersurface ({\it inertial observer}). The stress energy tensor is then
\beq
T^{\mu\nu}_{(\nu_i)} = E_{(\nu_i)} n^\mu n^\nu + F_{(\nu_i)}^\mu n^\nu + F_{(\nu_i)}^\nu n^\mu + P_{(\nu_i)}^{\mu\nu},
\eeq
with $F_{(\nu_i)}^\mu n_\mu = P_{(\nu_i)}^{\mu\nu}n_\mu = P_{(\nu_i)}^{\mu\nu}n_\nu = 0$. For convenience, in the rest of this paper we drop the subscript $(\nu_i)$ when referring to moments of $f_{(\nu_i)}$, but moments should always be understood as referring to a specific neutrino species. The moment formalism provides us with evolution equations for $E$ and $F_i$, the spatial components of $F_\mu$. They can be expressed in the familiar form
\beq
\nabla_\mu T^{\mu\nu}_{(\nu_i)} = Q^\nu_{(\nu_i)}
\label{eq:Trad}
\eeq
for some source terms $Q^\nu_{(\nu_i)}$ capturing interactions between neutrinos of species $(\nu_i)$ and the fluid, as well as interactions with other neutrino species. We also evolve the number density of neutrinos as measured by an inertial observer, $N$ (see~\cite{Foucart:2016rxm}). Finally, the evolution of the fluid is given by the equations
\beqn
\nabla_\mu T^{\mu\nu}_{\rm fl} &=& -\sum_{(\nu_i)} Q^\nu_{(\nu_i)},\\
\nabla_\mu \left(\rho \sqrt{-g} u^\mu\right) &=&0,
\eeqn
with $T^{\mu\nu}_{\rm fl} $ the stress-energy tensor of the fluid, $\rho$ the baryon density, and $g$ the determinant of the 4-metric. 

Eq.~(\ref{eq:Trad}) is exact, but depends on the unknown pressure tensor of the neutrinos, $P_{ij}$. In SpEC, we close the system of equations using the M1 closure~\cite{Minerbo1978}. Effectively, $P_{ij}$ is estimated by interpolating between its analytically known value in optically thick regions (isotropic pressure in thermal equilibrium with the fluid) and its value for a single beam of neutrinos propagating in vacuum. This is expected to be very accurate in regions of high optical depth, qualitatively correct in semi-transparent regions, and completely wrong in optically thin regions if neutrinos come from more than one direction. Once we have chosen a closure $P_{ij}(E,F_i)$, Eq.~(\ref{eq:Trad}) is a system of 4 equations for the 4 unknown $(E,F_i)$, for each neutrino species.

Besides a choice of closure, the gray two-moment scheme relies on significant approximations in the computation of $Q^\nu_{(\nu_i)}$. We include in $Q^\nu_{(\nu_i)}$ charged-current reactions
\beqn
p + e^-  &\leftrightarrow& n + \nu_e\\
n + e^+  &\leftrightarrow& p + \bar\nu_e,
\eeqn
$\nu\bar\nu$ pair annihilation/creation
\beq
e^+ + e^-  \leftrightarrow \nu + \bar\nu,
\eeq
plasmon decays
\beq
\gamma \leftrightarrow \nu + \bar\nu,
\eeq
and, for heavy-lepton neutrinos, nucleon-nucleon Bremsstrahlung (note that $\bar\nu_x = \nu_x$ in SpEC) 
\beq
N + N  \leftrightarrow N + N + \nu_x + \bar\nu_x.
\eeq
All of the emissivities and absorption opacities are computed following~\cite{Rosswog:2003rv}, except for nucleon-nucleon Bremsstrahlung~\cite{Burrows2006b}. We compute the neutrino absorption opacities $\kappa_{a,(\nu_i)}$ due to charged-current reactions and the neutrino emissivities $\eta_{(\nu_i)}$ due to other processes. The emissivities due to charged-current reactions and absorption opacities due to other processes are computed by imposing Kirchoff's law, $\eta = B \kappa_a$, with $B$ the energy density of neutrinos in equilibrium with the fluid. Using Kirchoff's law guarantees that we recover the correct neutrino energy density in optically thick regions. We also compute the scattering opacities $\kappa_{s,(\nu_i)}$ due to elastic scattering of neutrinos on neutrons, protons, and heavy nuclei~\cite{Rosswog:2003rv}, and estimate
\beq
Q^\nu_{(\nu_i)} = \eta_{(\nu_i)} u^\nu - \kappa_{a,(\nu_i)} J u^\nu - (\kappa_{a,(\nu_i)}+\kappa_{s,(\nu_i)}) H^\nu.
\eeq
We ignore inelastic scatterings, as well as all processes not explicitly mentioned here.

An important issue when computing $Q^\nu_{(\nu_i)}$ is that the cross-sections for the above reactions depend on the energy spectrum of the neutrinos. In a gray scheme, we can only guess at what that spectrum is. To compute opacities, we thus first assume that the neutrinos are in thermal equilibrium with the fluid. We then compute the average energy of the neutrinos from their moments $(E,F_i,N)$, and rely on a fairly complex and somewhat arbitrary procedure to estimate the shape of the neutrino spectrum in optically thin regions. We then correct the absorption and scattering opacities, assuming a $\nu^2$ dependence for the dominant neutrino-matter interactions (see~\cite{Foucart:2016rxm}). 

The situation is even worse in simulations that do not evolve the number density $N$. Then, even finding a good estimate of the average energy of neutrinos can be difficult. For neutron star mergers, this leads to large errors in the absorption rate of neutrinos in optically thin regions as well as in the composition of polar outflows~\cite{Foucart:2016rxm}.

Another problem with the two-moment scheme is that the rate of $\nu\bar\nu$ pair annihilation in optically thin regions is highly dependent on the momentum distribution of neutrinos, as the pair annihilation cross-section grows rapidly with the angle between the direction of propagation of the neutrino and the antineutrino. The pair annihilation rate is also very poorly approximated by Kirchoff's law in regions where the number density of electrons or positrons is low. We discuss below an approximate treatment of these annihilation processes, proposed by Fujibayashi et al.~\cite{Fujibayashi2017abc}. Within the framework of a gray two-moment scheme, however, any estimate of $\nu\bar\nu$ pair annihilation has potentially large errors.

To quantify these errors in the two-moment scheme, we rely on a newly developed MC scheme, described in detail in~\cite{Foucart:2017mbt}. The MC algorithm implemented in SpEC is largely inspired by earlier work on MC evolution of neutrinos in special relativity~\cite{richers:15} and on general relativistic photon transport~\cite{Ryan2015}. The MC scheme can theoretically be used as a closure for the two-moment algorithm. Doing this leads to a scheme that evolves Boltzmann's equations to numerical accuracy. For this first use in merger simulations, however, we consider a simpler, cheaper, and possibly more stable setup (the stability of the coupled M1-MC system has not been demonstrated). We evolve the equations of radiation-hydrodynamics using the two-moment scheme with M1 closure (we refer to this as the M1 scheme in the rest of this paper). The resulting time-dependent fluid quantities are used as background for the MC evolution. The MC evolution does not feed back onto either the fluid or the two-moment evolution. Each MC packet is created with a fluid frame energy of $10^{-11} M_\odot c^2$.

The MC algorithm implemented in SpEC is currently capable of handling isotropic emission of neutrinos in the fluid frame, transport of neutrinos along geodesics, neutrino absorption, and elastic scattering. We use the publicly available NuLib library~\cite{OConnor:2015} to generate a table of neutrino emissivities and absorption/scattering opacities as a function of fluid-frame neutrino energy $\nu$, fluid density $\rho$, fluid temperature $T$, and fluid electron fraction $Y_e$. The table has 12 energy bins spanning $\nu \in [0,150]\,{\rm MeV}$, with a logarithmic spacing between bins (except that the first 2 bins have a width of $4\,{\rm MeV}$). We also use 51 equally spaced bins for $Y_e\in [0.035,0.55]$, 82 logarithmically spaced bins for $\rho \in [10^6,3.16\times 10^{15}]\,{\rm g/cm}^3$, and 65 logarithmically spaced bins for $T\in [0.05,150]\,{\rm MeV}$. In between tabulated points, we interpolate the logarithm of the energy-dependent opacities $(\kappa_a,\kappa_s)$ linearly in $Y_e$ and logarithmically in the other variables. Following Richers et al.~\cite{richers:15}, we always emit particles with a fluid-frame energy at the center of an energy bin.
The NuLib table uses the same set of reactions as the moment scheme, except that it neglects all pair processes for $\nu_e \bar \nu_e$. The effective gray opacities derived from the MC evolution could however be very different from those in the moment scheme, as the MC algorithm is fully energy dependent while the moment scheme arbitrarily assumes a given neutrino spectrum at each point.

An important property of our MC algorithm is that it ignores regions of high optical depth, where the two-moment scheme is reliable and the neutrino distribution function is well approximated by a thermal distribution in equilibrium with the fluid. The MC algorithm only evolves regions of the post-merger remnant where $\kappa_a (\kappa_a + \kappa_s) \lesssim \kappa_{\rm crit}^2$. In any cell that does not satisfy this condition, but with a neighboring cell that does, MC particles are erased at the end of each time step and re-drawn from an equilibrium distribution. This provides a boundary condition for our MC algorithm. We note that we evaluate opacities separately for each energy bin, and that "neighbors" are determined in a 4D space (3 spatial dimensions, plus $\nu$). 

\begin{figure}
\begin{center}
\includegraphics[width=.99\columnwidth]{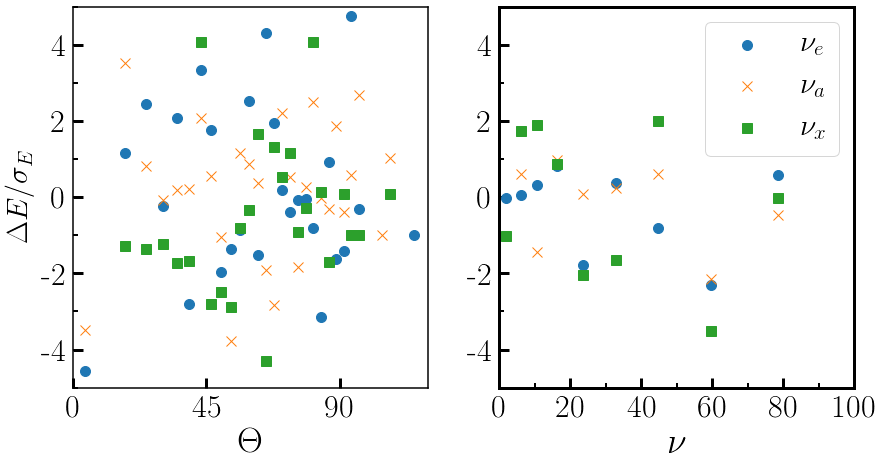}
\caption{Difference in the pitch-angle distribution and energy spectrum of neutrinos between two simulations placing the boundary of the MC evolution at different optical depth. We look at neutrinos within a $3\,{\rm km}$ radius of a point on the polar axis ($z=45\,{\rm km}$), about $5\,{\rm ms}$ after the beginning of the simulation. Errors are normalized by the expected Poisson error in our reference simulation. Both simulation have near-identical Poisson noise.}
\label{fig:varyK}
\end{center}
\end{figure}

We choose $\kappa_{\rm crit}$ so that in any region where the characteristic length scale for variation of the opacities is at least a few grid cells, the assumption of an equilibrium distribution of neutrinos in these ``high-opacity'' cells is reasonably accurate. For example, for an optically thick sphere of constant opacity $\kappa_0$, the relative difference between the energy density of neutrinos and the equilibrium density of neutrinos at distances $\Delta x$ from the surface is $\sim(0.08,0.02,0.005)$ for optical depth $\tau = \kappa_0 \Delta x = (1,2,3)$\footnote{We can also compute the ratio $|F|/E$, a measure of the anisotropy in the neutrino distribution function. For the same values of $\kappa_0 \Delta x$, we have $|F|/E\sim (0.056,0.015,0.005)$}. For this first simulation, we take $\kappa_{\rm crit} \Delta x \sim 1.2$. Regions where the opacities vary rapidly on the scale of a grid cell or less may be poorly approximated by our boundary condition, but are also underresolved in the two-moment scheme. A full MC evolution within these cells would not help much either, as our MC scheme assumes constant opacities and emissivities within any given cell. On the other hand, if the opacity varies on a length scale of more than 3 grid cells, our boundary condition should be accurate to better than $1\%$. 

To gain more confidence in this choice, we perform a shorter simulation with $\kappa_{\rm crit} \Delta x \sim 12$, and look at the neutrino distribution function above the neutron star ($z=45\,{\rm km}$). We choose this point because the polar cap of the neutron star has the steepest opacity gradients, and we thus expect polar regions to be particularly sensitive to a bad choice of $\kappa_{\rm crit}$. We find that differences between the two simulations for the flux of neutrinos, their energy spectrum, and their momentum distribution are close to the expected statistical noise, indicating that our boundary condition is not a dominant source of error at the accuracy currently reached by our code. Fig.~\ref{fig:varyK}, for example, shows deviations in the pitch-angle and energy distribution of the neutrinos, normalized to the expected Poisson noise of {\it one} of the simulations (both simulations have similar Poisson noise). For an absolute error scale, we note that this figure was generated using $\sim (4300,7400,2200)$ packets per simulation for $(\nu_e,\bar\nu_e,\nu_x)$. We also measured differences in the average energy of the neutrinos of $\Delta \langle \epsilon \rangle = (0.15,0.01,0.7)\,{\rm MeV}$ between the two simulations.

Our analysis of the MC results relies on two types of data. First, we have at our disposal the individual packets evolved by the MC scheme, and we log information about all MC packets leaving the computational domain. This allows us to obtain a MC estimate of the distribution function at any given point, as long as we compute it on-the-fly, and to post-process at will information about the neutrinos leaving the grid. Second, we compute time-averages of moments of the neutrino distribution function. These are meant to be used, eventually, to provide a better closure to the two-moment scheme~\cite{Foucart:2017mbt}. In this paper, they allow us to compare the M1 and MC results, and they also serve in the computation of the $\nu\bar\nu$ pair-annihilation rate. We use time-averaged moments so that a lower number of MC packets can be used in the simulation. This significantly reduces computational costs: in this study, the MC evolution is actually cheaper than the M1 evolution. With $\kappa_{\rm crit} \Delta x \sim 1.2$ and each MC packet having an energy of $10^{-11}M_\odot c^2$, we have roughly as many MC packets on the grid as we have finite volume grid cells. All moments are computed by averaging over 100 MC packets (or one time-step if more than 100 packets are present in a cell)\footnote{The exact procedure to compute time-averaged moments is described in~\cite{Foucart:2017mbt}}, leading to expected relative errors from Poisson noise of $\lesssim 10\%$.

\subsection{Post-merger initial conditions}

\begin{figure*}
\begin{center}
\includegraphics[width=.99\textwidth]{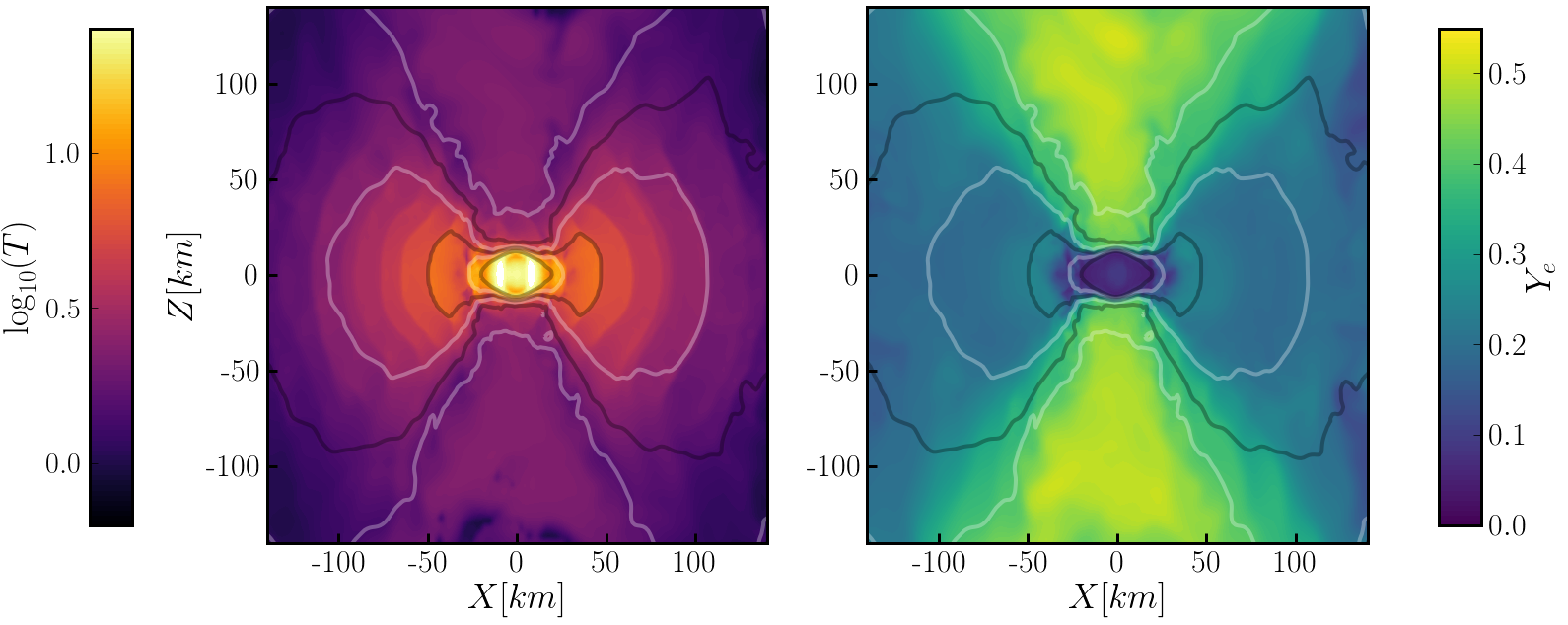}\\
\includegraphics[width=.99\textwidth]{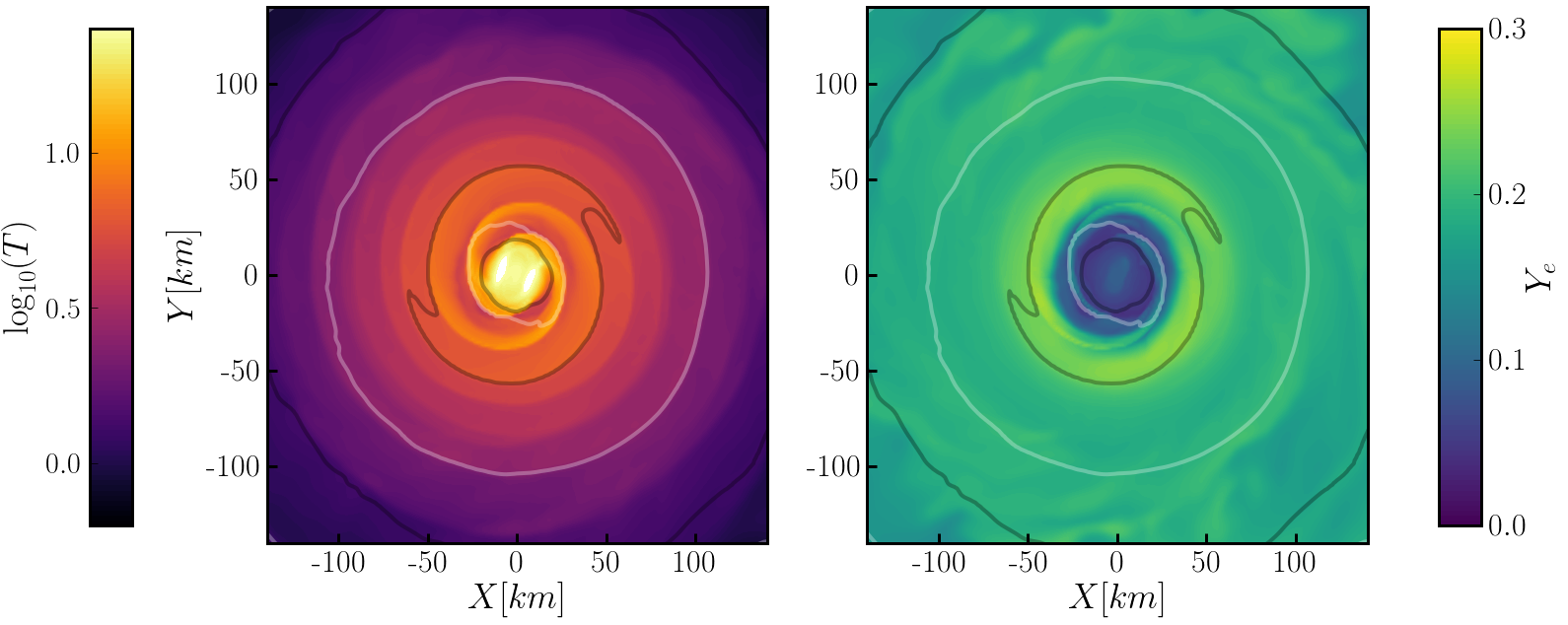}
\caption{Poloidal (top) and equatorial (bottom) slices through the merger remnant. In each figure, solid lines show density contours of
$\log_{10}(\rho/[1\,{\rm g/cm^3}])=(8,9,10,11,12,13,14)$. Color scales show the fluid temperature (left) and electron fraction (right).}
\label{fig:FluidSlices}
\end{center}
\end{figure*}

We choose as initial conditions the result of one of our existing simulation of merging neutron stars~\cite{Foucart:2016rxm}. The initial binary is an equal mass, non spinning system. Each neutron star has an ADM mass of $1.2M_\odot$ in isolation. The neutron star matter is described by the equation of state of Lattimer \& Swesty~\cite{Lattimer:1991nc} with nuclear incompressibility parameter
$K_0 = 220\,{\rm MeV}$ (LS220). We use the publicly available NuLib table providing the fluid properties as a function of density ($\rho$), temperature ($T$), and electron fraction ($Y_e$) for this equation of state~\cite{OConnor2010}. In this work, we begin from a snapshot of the simulation $10{\rm ms}$ after merger, and evolve the post-merger remnant for $4.5\,{\rm ms}$. This is longer than the dynamical timescale of the remnant neutron star, but shorter than the cooling timescale of the remnant. 

The fluid properties within poloidal and equatorial slices through the remnant are shown in Fig.~\ref{fig:FluidSlices}. The central object is a hot, differentially rotating, massive neutron star. It is surrounded by a thick accretion disk with temperature $T\sim 5-10\,{\rm MeV}$ and electron fraction $Y_e \sim 0.1-0.3$.
Shocked spiral arms driven by the rotation of the distorted neutron star are visible in the disk. Low-density outflows are observed in the polar regions. Neutrino emission and absorption make these outflows relatively proton-rich ($Y_e \sim 0.3-0.5$). In previous work, we demonstrated that the composition of these outflows is quite sensitive to the method used to compute the average energy of the neutrinos~\cite{Foucart:2016rxm}. The starting point of this study is a simulation using our best energy estimate so far for neutrino energies, i.e. the estimate obtained by evolving both the neutrino number density and the neutrino energy density. 

Our existing simulation provides us with initial conditions for the metric, the properties of the fluid, and the moments of the neutrino distribution function evolved by the M1 scheme. These have to be complemented with initial conditions for the MC evolution. At the initial time, we randomly draw particles from a thermal distribution in equilibrium with the fluid. While this is a fairly reasonable assumptions in the most optically thick regions evolved by the MC algorithm, this is clearly inexact in semi-transparent and optically thin regions. The duration of the simulation is chosen to allow MC packets to diffuse from the surface of moderate optical depth below which we do not use the MC algorithm (discussed in the previous section), and then travel to the boundary of the domain. In the rest of this paper, we largely ignore the first $\sim 3\,{\rm ms}$ of evolution, and focus solely on times when we expect that the properties of the MC packets are no longer influenced by our choice of initial conditions.

\subsection{Numerical grids}

We set our numerical grids as in~\cite{Foucart:2016rxm}. The pseudo-spectral grid is constructed from a small filled sphere centered on the neutron star remnant, surrounded by 59 spherical shells. The number of basis functions within each subdomain is chosen so that the truncation error in the metric and in its spatial derivatives is less than $5 \times 10^{-4}$. Given the near-spherical symmetry of the high-density regions of the post-merger remnant, this requires only $\sim 140k$ grid points. 

The finite volume grid is constructed from 4 levels of refinement. The finest level has a grid spacing $\Delta x\sim 300\,{\rm m}$. At each subsequent level, the grid spacing is multiplied by 2. All levels have the same number of grid cells, $200\times200\times100$, and are centered on the neutron star remnant. Each level is further subdivided in $144$ patches, for parallelization. Each patch is then extended by 3 {\it ghost cells} in each direction, for reconstruction of the fluid variables from grid centers to faces. Overall, the grid has $576$ patches and a total of $\sim 28$ million cells. As we aim to estimate differences between the MC and M1 schemes for a fixed setup and at a limited computational cost, we do not attempt to vary the grid spacing or test the convergence of the fluid evolution. Our grid spacing is fairly typical for neutron star merger simulations with SpEC, and can capture the dynamics of the post-merger remnant~\cite{Foucart:2016rxm}. Our grid would, on the other hand, be far too coarse to study the effects of magnetic fields~\cite{Kiuchi2015}, which are entirely ignored in this work.

\section{Neutrino moments and distribution function}

\subsection{Eddington tensor}

\begin{figure*}
\begin{center}
\includegraphics[width=.99\textwidth]{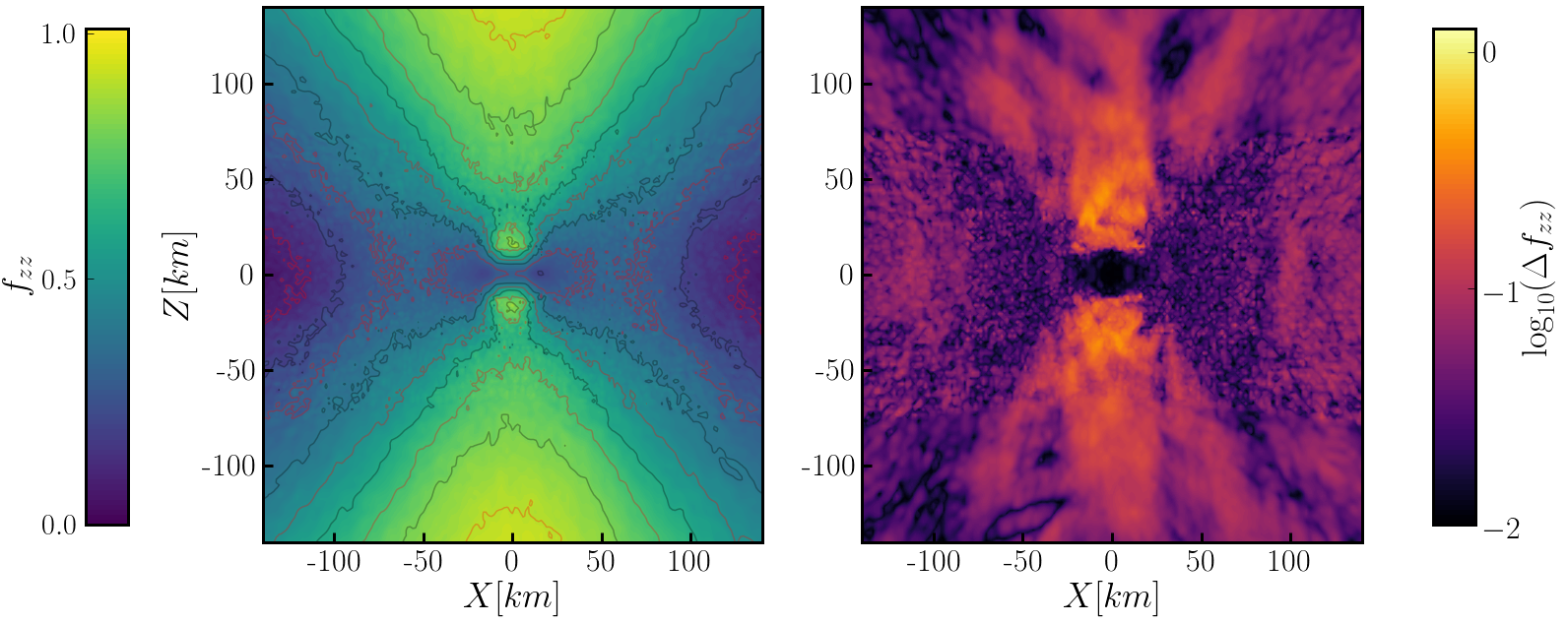}\\
\includegraphics[width=.99\textwidth]{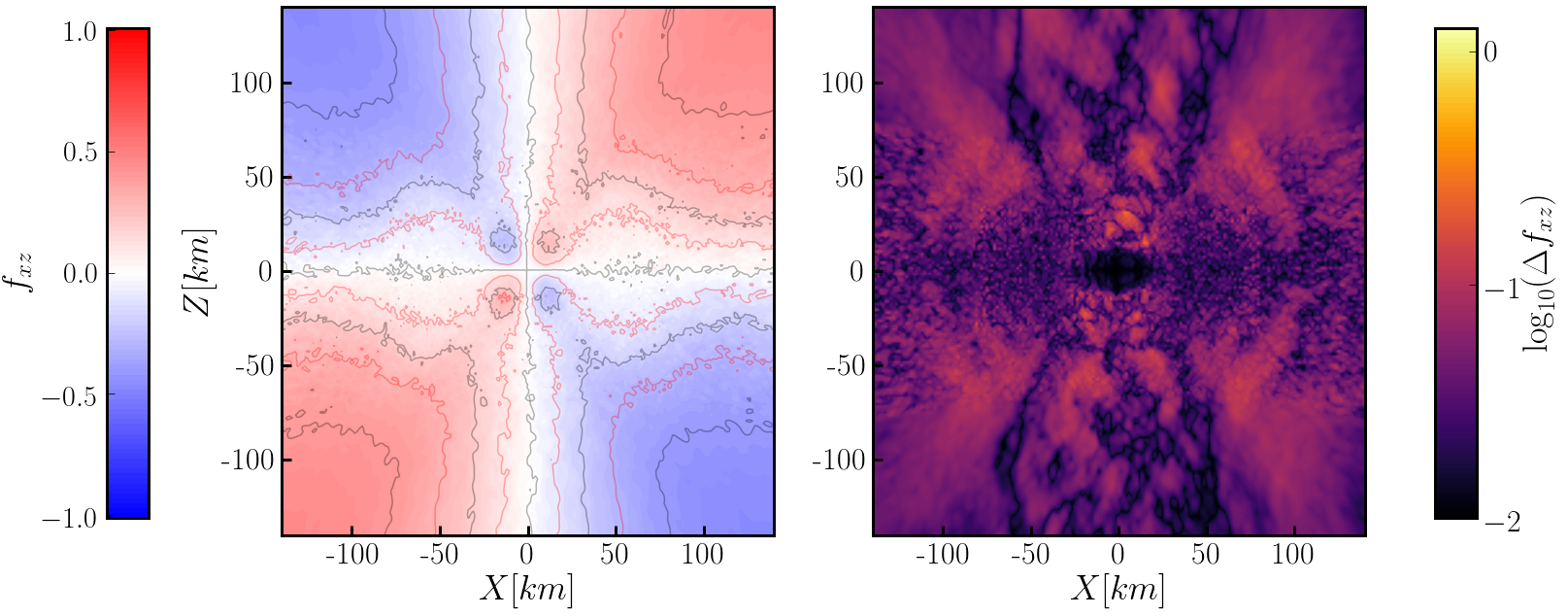}
\caption{Poloidal slice through the merger remnant. The left column shows the diagonal and off-diagonal components of the Eddington tensor $f_{zz}$ and $f_{xz}$ as measured with the MC code. Solid lines are contours of $f_{zz},f_{xz}$ separated by $0.1$ (alternating red and black lines). This provides an order-of-magnitude estimate of the statistical noise, with $\Delta f_{\rm stat} < 0.05$.
The right column shows the difference between the MC and M1 Eddington tensors. In the massive neutron star, the difference is negligible. In the disk, it is dominated by statistical MC noise. Outside of the remnant, large errors in the M1 closure dominate. In particular, $\Delta f_{zz}\sim 0.1-0.3$ in the polar regions. All results are for $\nu_e$ neutrinos.}
\label{fig:EddingtonVert}
\end{center}
\end{figure*}

\begin{figure*}
\begin{center}
\includegraphics[width=.99\textwidth]{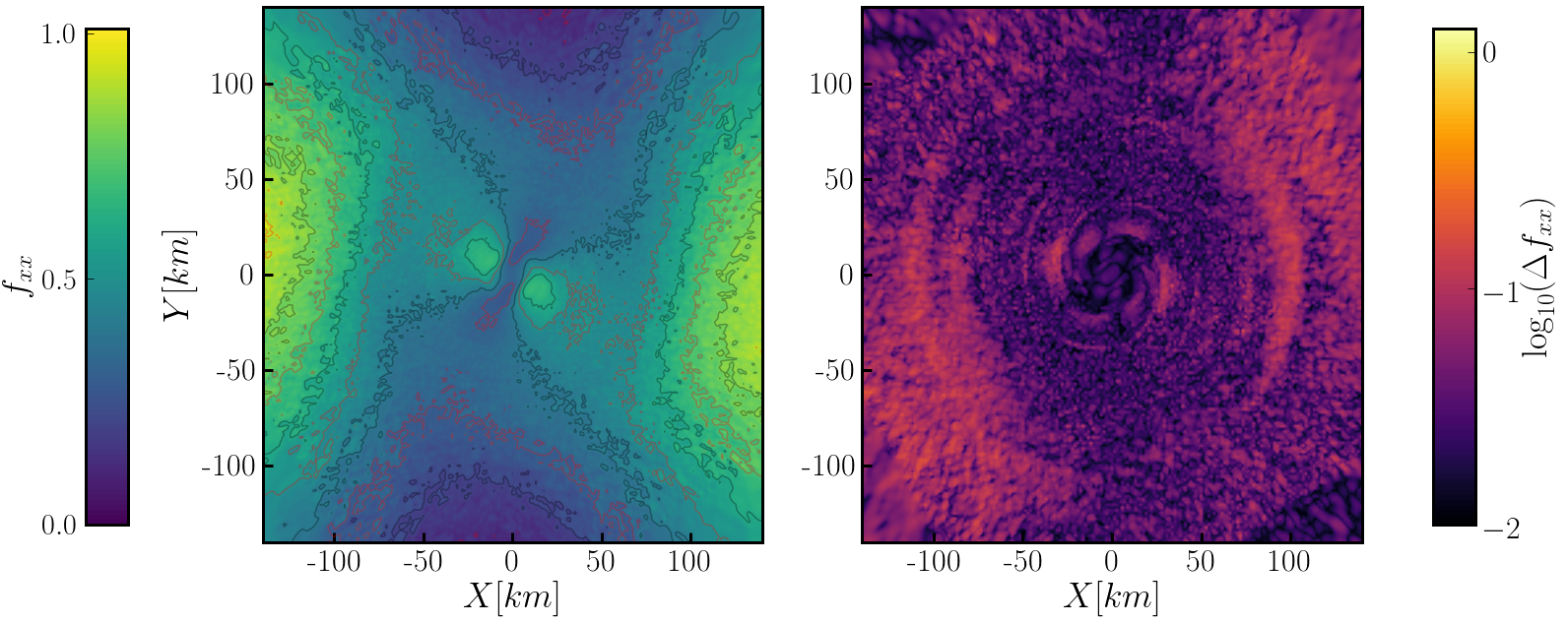}\\
\includegraphics[width=.99\textwidth]{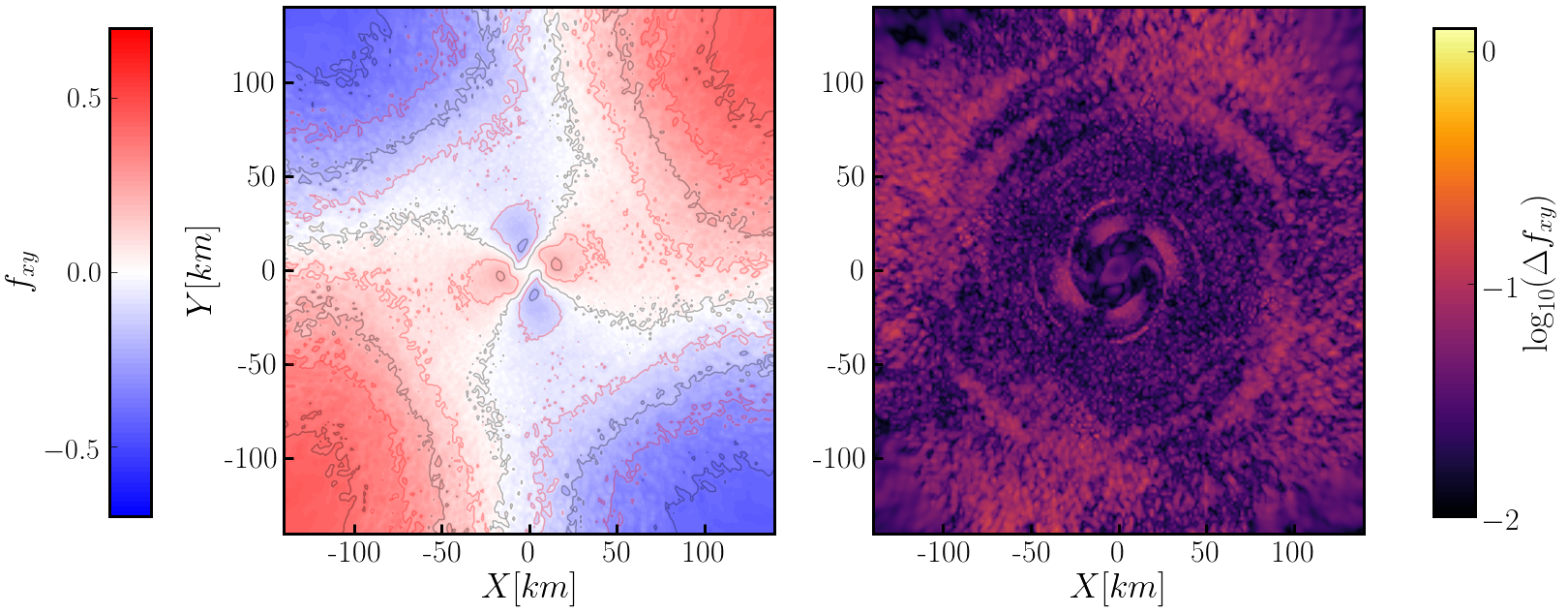}
\caption{Same as Fig.~\ref{fig:EddingtonVert}, but for an equatorial slice through the remnant, and showing components $f_{xx},f_{xy}$ of the Eddington tensor.
The M1 closure is significantly more accurate here than in the polar regions, although some regions of the shocked spiral arms show consistent biases in the M1 closure at a level $\Delta f_{ij} \sim 0.1$.}
\label{fig:EddingtonHor}
\end{center}
\end{figure*}

We begin our analysis by considering the impact of one of the main assumption of the two-moment scheme: the M1 analytical closure. In Figs.~\ref{fig:EddingtonVert}-\ref{fig:EddingtonHor}, we show equatorial and poloidal slices through the merger remnant, as for the fluid quantities plotted in Fig.~\ref{fig:FluidSlices}. The figures show various components of the Eddington tensor $f_{ij} = P_{ij}/E$, in an orthonormal frame constructed by an inertial observer applying the Gramm-Schmidt algorithm to vectors tangent to spatial coordinate lines. For each component we plot the MC results, and the difference between MC and M1 results. From the MC results, we see that statistical errors, roughly approximated as the noise in the MC predictions, are at the level of a few percents, slightly better than expected.

The difference between MC and M1 is largest in the polar regions, and for $f_{zz}$. Errors of $0.1-0.3$ are the norm within a few neutron star radii of the surface of the remnant, with the M1 closure consistently returning larger values of $f_{zz}$ than the MC closure. That the M1 closure is particularly inaccurate in polar regions is no surprise: neutrinos emitted by the hot neutron star and the accretion disk cross paths there, and will create artificial radiation shocks when the M1 closure is used. With this simulation, we can quantify this long-standing assumption. We find that errors in the polar regions are very significant: the MC results indicate that $f_{zz}\sim 0.5-0.7$ at points where the difference between MC and M1 is $|\Delta f_{zz}| \sim 0.1-0.3$. Outside of the polar regions, we observe differences $|\Delta f_{ij}| \lesssim 0.1$. In some of these regions, the difference between the MC and M1 results is consistently of the same sign, and thus does not appear due to statistical noise in the MC results. Regions with rapid variations in the error measurements (with typically $\Delta f_{ij} \lesssim 0.03$), on the other hand, most likely have larger MC errors than M1 errors. This is the case in most of the accretion disk, at least at radii $r\lesssim 70\,{\rm km}$.

Overall, the M1 closure appears to do quite well in the optically thick and semi-transparent regions where most of the neutrinos are emitted, but has some clear issues farther from the remnant, where we pay for the inaccuracies of the optically-thin M1 closure. Large errors for the M1 closure in the polar regions have a couple of potentially important consequences for neutrino-matter interactions in these systems. One is that the spatial distribution of neutrinos in optically thin regions is inaccurate when using a M1 scheme. This impacts the resulting rate of absorption of $\nu_e$ and $\bar\nu_e$, and thus the evolution of the composition of polar outflows. We study the spatial distribution of neutrinos in more detail in Sec.~\ref{sec:nuang}. Another consequence is that the inferred energy deposition from $\nu\bar\nu$ annihilations in polar regions may be difficult to accurately estimate when using the M1 closure. We consider that problem in Sec.~
\ref{sec:pair}.

\subsection{Pointwise distribution function of neutrinos}

To better understand the momentum-space distribution of neutrinos, we now look at their direction of propagation and their energy spectrum at individual points. As opposed to our computation of the Eddington tensor, we do not perform any time-averaging here. Instead, we study the properties of all MC packets within a distance $\Delta d$ of a point $\bf{x}_i$ at a fixed time $t_i$. We also limit ourselves to optically thin regions, where differences between the M1 and MC results are significant. As the distribution function of neutrinos at a given time is 6-dimensional, and different physical processes will require the visualization of that distribution function in different ways, we do not attempt to provide a complete view of the distribution function. We limit ourselves to some notable properties of the distribution function, provided as examples of what information can be gleaned from our MC results. 

\begin{figure}
\begin{center}
\includegraphics[width=.99\columnwidth]{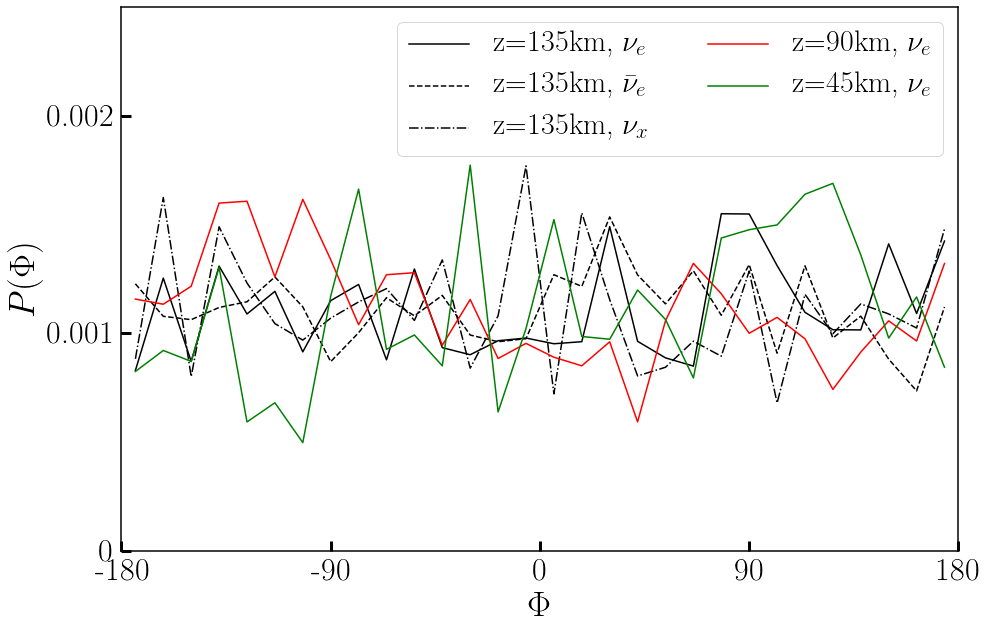}
\caption{Distribution probability of neutrinos as a function of the azimuthal angle $\phi$ of their 4-momentum, at different points on the polar axis. We sample all packets within $\Delta d = 0.1z$ of the target point. We have $400-2000$ packets within each region and for each type of neutrinos.}
\label{fig:phidis}
\end{center}
\end{figure}

At each point, the momentum of neutrinos is characterized by the parameters $(\epsilon,\theta,\phi$), where $\epsilon$ is the energy of the neutrinos measured by an inertial observer, $\theta$ the angle between the momentum of the neutrinos and the radial coordinate in an orthonormal tetrad constructed by an inertial observer (pitch-angle), and $\phi$ an azimuthal angle for rotation around that same radial axis. 
We first consider points along the polar axis, at $z=(45,90,135)\,{\rm km}$. At all 3 points, we find a flat distribution in $\phi$, within statistical errors and after marginalizing over $(\epsilon,\theta)$. This indicates that deviations from axisymmetry in the hot neutron star and in the surrounding accretion disk do not have a significant impact on the neutrino distribution function at the poles. This result is particularly interesting if we aim to use time-averaged MC moments to close the two-moment evolution equations. It may be sufficient for the averaging timescale to be short compared to the thermal evolution timescale of the remnant, rather than its orbital timescale.

\begin{figure}
\begin{center}
\includegraphics[width=.99\columnwidth]{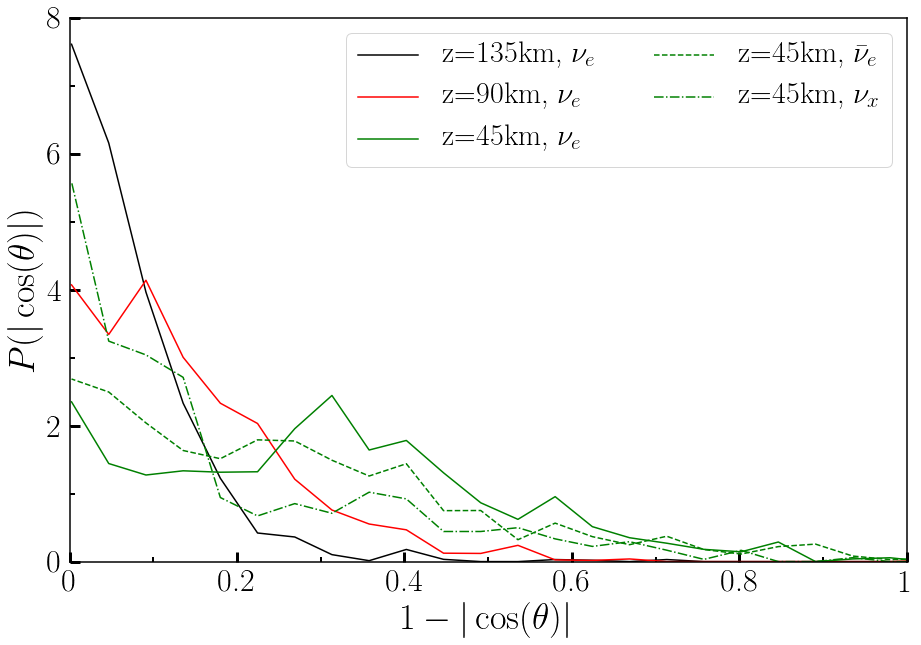}
\caption{Same as Fig.~\ref{fig:phidis}, but for the distribution probability with respect to $\cos(\theta)$, with $\theta$ the neutrino pitch angle. We show multiple species for the closest point to the remnant, as the angular distribution is more sensitive to the finite size of the emitting region close to the remnant.}
\label{fig:thetadis}
\end{center}
\end{figure}

Fig.~\ref{fig:thetadis} shows the probability distribution of neutrinos with respect to $\theta$. As expected, it becomes more forward-peaked as we move away from the remnant. We can also see clear differences between neutrino species. The pitch-angle distribution is narrower for the heavy-lepton neutrinos, and wider for $\bar\nu_e$. For $\nu_e$, the distribution function peaks at $\theta \sim 45^o$ rather than $\theta\sim 0^o$. This is due to the relative contribution of the neutron star and accretion disk to the neutrino fluxes. Nearly all $\nu_x$ are coming from the neutron star, while the disk contributes significantly to the production of $\bar\nu_e$ and $\nu_e$. The broad distribution of $f(\theta)$ does not match the assumptions made by the M1 closure in these regions.

\begin{figure}
\begin{center}
\includegraphics[width=.99\columnwidth]{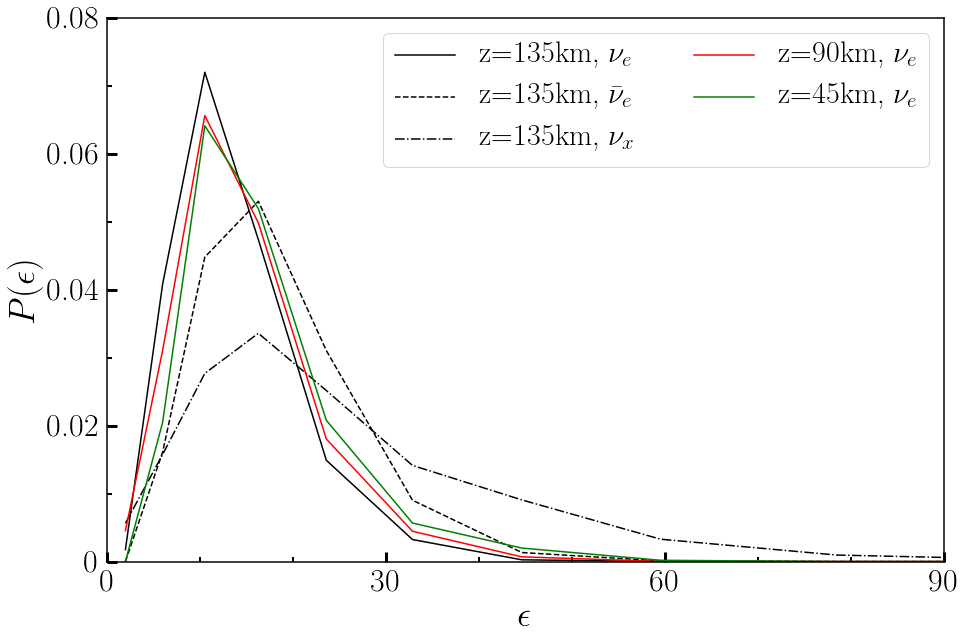}
\caption{Same as Fig.~\ref{fig:phidis}, but for the distribution probability with respect to the neutrino energy $\epsilon$. We use 12 energy bins to generate this figure, identical to the bins of the NuLib table.}
\label{fig:endis}
\end{center}
\end{figure}

Finally, the energy spectrum of neutrinos is shown in Fig.~\ref{fig:endis}. The shift in the spectrum as we move away from the remnant is too large to be due to gravitational redshift alone. We expect a $\sim 6\%$ shift in the average energy of the neutrinos between $z=45\,{\rm km}$ and $z=135\,{\rm km}$, but observe a $20\%$ change between those points. The cooler spectrum at large radii is a geometrical effect, accounting for a larger fraction of the polar neutrinos coming from the disk rather than the hotter neutron star as we move away from the remnant along the polar axis. The spectra also show that the assumption of a black-body (or softer) spectrum made in the gray moment scheme~\cite{Foucart:2016rxm} is not accurate. We fit the normalized spectra at $z=135\,{\rm km}$ to the function
\beq
f(\epsilon,T_{(\nu)},\alpha) = \frac{\epsilon^\alpha}{T_{(\nu)}^{\alpha+1}\Gamma(\alpha-1)}e^{-\epsilon/T_{(\nu)}}.
\label{eq:ffit}
\eeq
For a black body spectrum, and approximating the Fermi-Dirac distribution of neutrinos by a Boltzmann distribution, we would expect $\alpha=2$ and $T_{(\nu)}$ would be the temperature of the neutrinos. Instead, for $(\nu_e,\bar\nu_e,\nu_x)$ we find $\alpha=(4.2,4.3,1.7)$ and $T_{(\nu)}=(2.5,3.3,9)\,{\rm MeV}$. The neutrino spectrum is thus significantly harder than expected for $\nu_e$ and $\bar\nu_e$, and closer to a black body for heavy-lepton neutrinos. The average energy of polar neutrinos evaluated from the gray moments $(E,F_i,N)$, on the other hand, is quite close to the average energy measured in the MC code: the average energy of neutrinos leaving the grid with a momentum misaligned by less than $30^0$ with respect to the polar axis, as measured in the moment scheme, is within $10\%$ of the MC results.

Given the expected dependence of the absorption and scattering opacities of neutrinos in the square of the neutrino energies, we can estimate that these different spectral shape would lead us to underestimate reaction rates for $\nu_e$ and $\bar\nu_e$ by up to $\sim 15\%$, if the average energy of neutrinos was accurately estimated. As the average energy of the polar neutrinos is within $10\%$ of the MC results in the M1 scheme, absorption and scattering opacities in the polar regions will be accurate within $\sim 30\%$. We note that this is only true because we evolve the neutrino number density in the M1 scheme, thus obtaining a reasonably accurate local estimate of the average neutrino energy. If we had approximated the neutrino energy spectrum by a black-body distribution at the fluid temperature ($T_{(\nu)}\sim (2-3)\,{\rm MeV}$, $\alpha=2$), the average neutrino energies would have been off by factors of $2-5$. 

We also study the properties of the neutrino distribution function at points farther away from the polar axis, in the $y=0$ plane (i.e. the poloidal slice shown in Fig.~\ref{fig:FluidSlices}). We consider the points 
\beqn
(x,z)&=&[(45,135),(90,135),(135,135),(135,90), \nonumber \\
&&(135,45),(135,0)]\,{\rm km}.
\eeqn
While at those points the probability distribution with respect to the angle $\phi$ is no longer isotropic, the only asymmetries observed in the neutrino distribution function are the expected preference for neutrinos to come from the equatorial plane of the remnant, and a more forward-peaked distribution function at larger distances. This is consistent with what we observed in the polar regions, and with our assumption that the neutrino distribution function does not vary significantly over the rotation period of the remnant. The spectrum of the neutrinos remains well fitted with the same function as in the polar regions, with a hard spectrum ($\alpha \in [4,5]$) for $\nu_e$ and $\bar\nu_e$ and a near black-body spectrum for $\nu_x$. The accuracy of the M1 results for $\epsilon$ decreases significantly as we move away from the polar axis. We find relative errors of $(10-30)\%$ in the average neutrino energies, which could lead to close to a factor of 2 errors in absorption/scattering opacities. In these regions, however, the composition of the fluid is not as sensitive to estimates of $\epsilon$ as closer to the poles~\cite{Foucart:2016rxm}. The impact of this error on EM observables is thus likely to be reasonably small, compared to other existing simulation errors.

Overall, we estimate that the error in the scattering and absorption opacities computed in the two-moment scheme are likely $\lesssim 30\%$ in the polar regions, where they impact EM observables the most. Errors are larger farther away from the poles, but this may not matter as much for modeling EM signals. However, computing the opacities is only one part of the problem. The energy density of neutrinos also impacts the reaction rate for neutrino-matter interactions. We show in Sec.~\ref{sec:nuang} that this is a more significant issue.

\subsection{Properties of escaping neutrinos}
\label{sec:nuang}

\begin{figure*}
\begin{center}
\includegraphics[width=.99\textwidth]{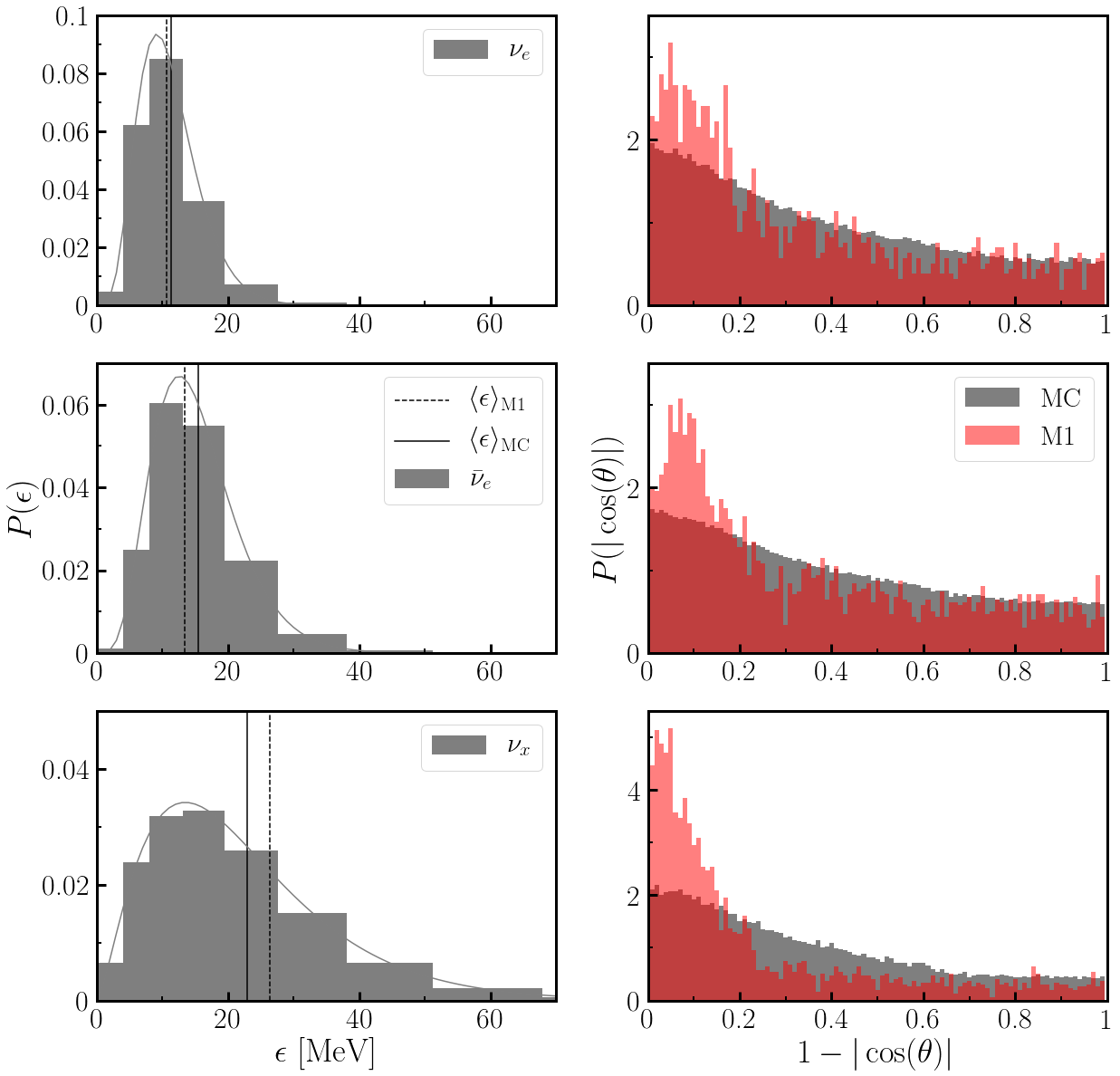}
\caption{{\it Left}: Energy distribution of the neutrinos leaving the computational domain, for all 3 species of neutrinos. In each plot, the dashed vertical line shows the average neutrino energy estimated by the M1 scheme, and the solid vertical line the same quantity estimated by the MC scheme. The solid grey line shows our best fit to the spectrum. {\it Right}: Angular distribution of the neutrinos leaving the grid. Here, $\theta$ is the usual spherical-polar coordinate, not the pitch angle of the neutrinos. Grey histograms show the MC results, and red histograms the M1 results. Errors in the M1 closure lead to a large overestimate of the neutrino density in the polar regions. In all plots, we integrate the neutrino fluxes over a $50\mu s$ interval $14{\rm ms}$ after merger.}
\label{fig:GlobalDis}
\end{center}
\end{figure*}

We now consider the properties of the neutrinos leaving the computational domain, starting with their energy spectrum. This illustrates differences between the energy of the neutrinos in the M1 and MC schemes, already discussed in the previous section. The spectrum of escaping neutrinos $14\,{\rm ms}$ after merger is shown on the left panel of Fig.~\ref{fig:GlobalDis}, for all three types of neutrinos. We bin the spectrum using the same 12 energy bins as in the NuLib table. It is worth noting, however, that our ability to resolve the energy distribution of neutrinos is understated in this plot. All MC packets are emitted with the energy of the center of a bin, but their energies can then be shifted due to gravitational and velocity redshift as well as scattering events, so that for example a global shift of the spectrum by a fraction of an MeV would be captured by the MC code.  

As for the pointwise data, the energy spectrum of $\nu_e$ and $\bar\nu_e$ is well fitted by Eq.~(\ref{eq:ffit}) with a hard spectral index ($\alpha=4.8$ and $T_{(\nu)}=1.9\,{\rm MeV}$ for $\nu_e$;  $\alpha=4.6$ and $T_{(\nu)}=2.7\,{\rm MeV}$ for $\bar\nu_e$). The spectrum of heavy lepton neutrinos is slightly softer than a black body ($\alpha=1.5$, $T_{(\nu)}=9.0\,{\rm MeV}$). The average energy of escaping neutrinos is reasonably well estimated in the moment scheme for $\nu_e$ and $\bar\nu_e$ ($1-2\,{\rm MeV}$ errors), while larger errors are observed for $\nu_x$ ($4\,{\rm MeV}$).

The right panel of Fig.~\ref{fig:GlobalDis} is more significant. There, we show the probability distribution of neutrinos as a function of their latitude in a spherical polar coordinate system with axis aligned with the angular momentum of the remnant. In this case, we see very significant differences between the MC and M1 results. Artificial shocks cause neutrinos to accumulate close to the polar axis in the M1 code, an effect that is avoided in the MC code. This results in an excess of neutrinos in the polar regions ($\theta\lesssim 30^0$), by $\sim 50\%$ for $\nu_e$ and $\bar\nu_e$ and by nearly a factor of 2 for $\nu_x$. This excess is likely to have a more important impact on the evolution of the polar outflows than the other sources of errors considered so far. The absorption rate of all flavors of neutrinos is significantly boosted in M1 simulations, causing excess heating of the outflows. The effect of this error on the composition of the outflows is hard to determine with certainty. Generally speaking, overestimated neutrino-matter reaction rates and fluid temperatures are likely to lead us to overestimate the electron fraction of the outflows in the M1 code. Quantifying this error will require simulations in which the MC scheme is fully coupled to the moment evolution (or directly to the fluid). We should note, however, that the magnitude of this effect may be overestimated by looking at Fig.~\ref{fig:GlobalDis}: most of the mass in the polar outflows is closer to $\theta\sim 30^o$ than to the polar axis, and the M1 closure performs better close to the disk/outflow boundary than on the polar axis.

\section{$\nu\bar\nu$ pair annihilation}
\label{sec:pair}

\begin{figure*}
\begin{center}
\includegraphics[width=.99\textwidth]{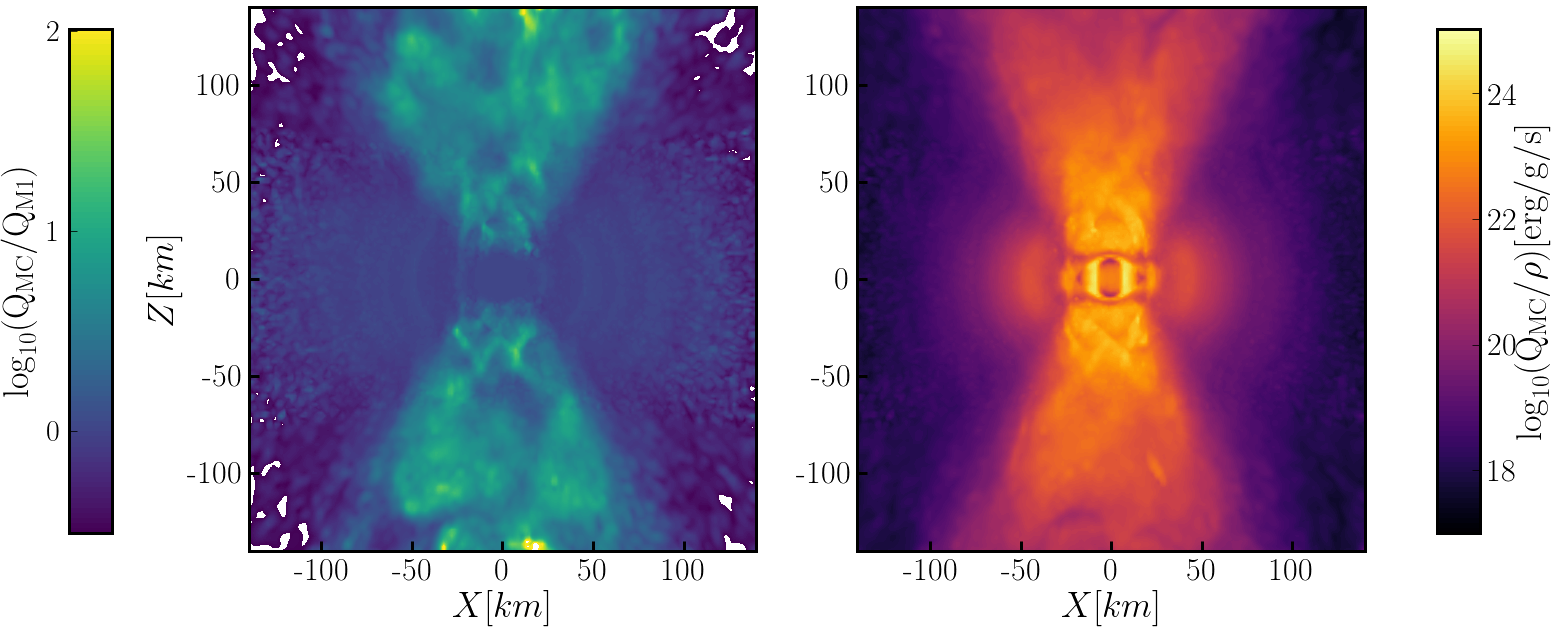}\\
\includegraphics[width=.99\textwidth]{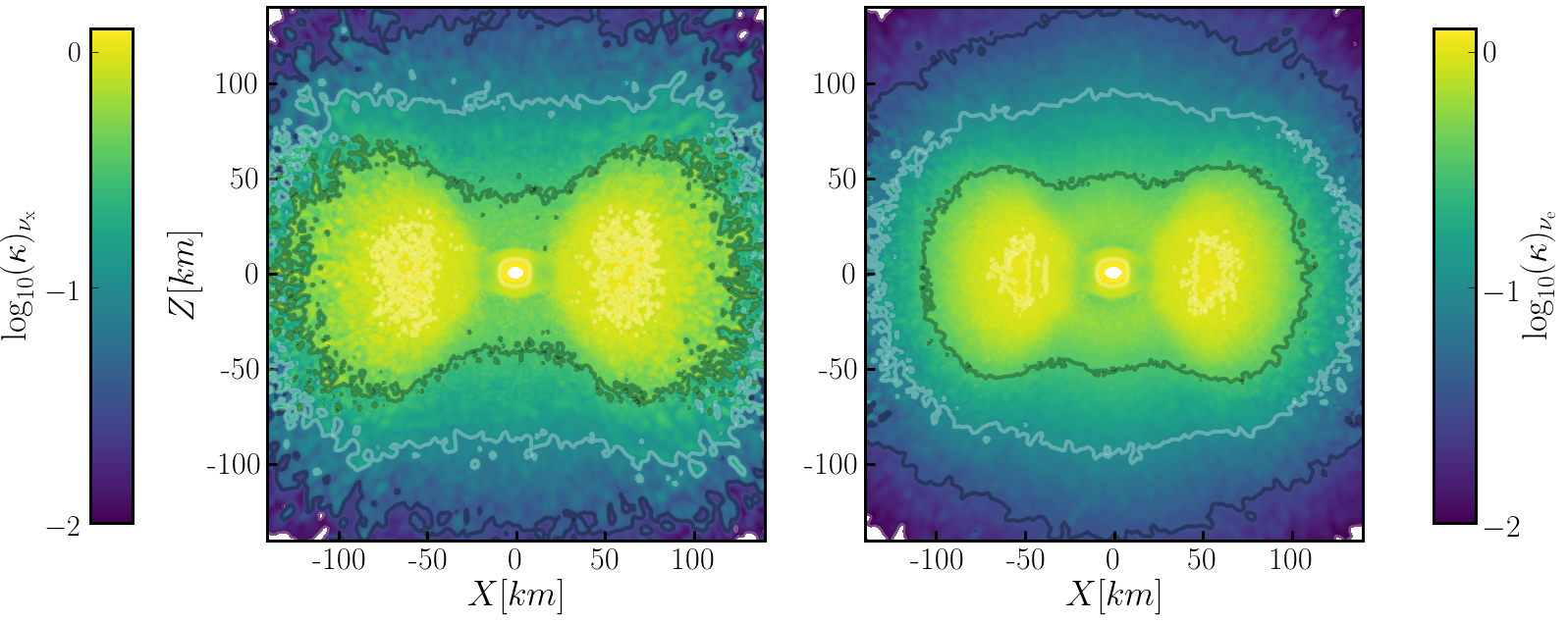}
\caption{Vertical slices through the merger remnant. Top: The specific energy deposition rate due to $\nu\bar\nu$ pair annihilation using the value of $\kappa$ predicted by the MC code (right), and the ratio of the predictions from the MC and M1 code (left). In both panels, we assume the neutrino energy density measured in the M1 code and the energy spectrum from the MC code. This provides an estimate of the error due solely to the use of the M1 closure in the computation of Eq.~\ref{eq:kappa}. Bottom: The geometric factor $\kappa$ in Eq.~\ref{eq:kappa}, for $\nu_x\bar\nu_x$ annihilation (left) and $\nu_e\bar\nu_e$ annihilation (right).}
\label{fig:PairAnnihilation}
\end{center}
\end{figure*}

So far, we have discussed aspects of neutrino transport and neutrino-matter interactions that are approximately modeled in M1 simulations. We now move to a potentially important physical effect that is entirely ignored in our existing M1 simulations: $\nu \bar\nu \rightarrow e^+ e^-$ pair annihilation in low-density polar regions. Existing estimates indicate that pair annihilation can deposit enough energy in the polar regions to drive mildly relativistic outflows, and clear the poles of most baryons -- although on their own they are probably not sufficient to power anything but the weakest short gamma-ray bursts~\cite{1999ApJ...517..859S,Birkl2007,just:16,Fujibayashi:2017puw,2017JPhG...44h4007P}. Pair annihilation has been included in 2D post-merger simulations using the two-moment approximation~\cite{just:16,Fujibayashi:2017puw}, and by post-processing late-time snapshots of a post-merger remnant~\cite{2017JPhG...44h4007P}, but not in self-consistent 3D simulations of these systems. One reason is that energy deposition due to pair annihilation is strongly dependent on moments of the neutrino distribution function that are not evolved by the M1 scheme, mainly because counter-propagating neutrinos have a much higher annihilation cross-section than neutrinos propagating in the same direction.

To study this effect, let us follow Fujibayashi et al.~\cite{Fujibayashi:2017puw} and assume that the phase-space blocking factors and masses of electrons and positrons are negligible. This is nearly certainly a good approximation in the low-density polar regions where  pair annihilation plays an important role. The heating rate $Q^{(+)}_{{\rm pair},\nu_i}$ due to the neutrino species $\nu_i$ can then be computed as a function of the moments of the neutrino distribution function through an integral over the phase space of neutrinos and antineutrinos (see~\cite{1999ApJ...517..859S,Fujibayashi:2017puw}). If the energy in the fluid frame $\omega_{(\nu_i)}$ can be factored from this integral and approximated by the average energy $\langle  \omega_{(\nu_i)} \rangle_{\rm pair}$ of neutrinos $\nu_i$, then
\beq
Q^{(+)}_{{\rm pair},\nu_i} = \frac{C^{\rm pair}_{\nu_i\bar\nu_i} G_F^2}{3\pi \bar h^4 c^3} \langle \omega_{(\nu_i)} \rangle_{\rm pair} (J \bar J - 2 H^\mu \bar H_\mu + S^{\mu\nu}\bar S_{\mu\nu})
\label{eq:Qpair}
\eeq
with $G_F \approx 4.5438 (\bar h c)^3 {\rm erg}^{-2}$ the Fermi constant,  ($\bar J,\bar H^\mu,\bar S^{\mu\nu})$ fluid-frame moments of antineutrinos,
\beq
C_{\rm pair}=1\pm 4 \sin^2(\theta_W) + 8 \sin^4(\theta_W)
\eeq
(with the plus sign for electron neutrinos and the minus signs for heavy-lepton neutrinos), and $\sin^2 \theta_W \approx 0.2319$. The total heating rate due to pair annihilation is then 
\beq
Q^{(+)}_{{\rm pair,tot}} = Q^{(+)}_{{\rm pair},\nu_e} + Q^{(+)}_{{\rm pair},\bar\nu_e} + 4 Q^{(+)}_{{\rm pair},\nu_x},
\eeq
where the only difference between $Q^{(+)}_{{\rm pair},\nu_e}$ and $Q^{(+)}_{{\rm pair},\bar\nu_e}$ is due to different average energies $\langle  \omega_{(\nu_i)} \rangle_{\rm pair}$ for $\nu_e$ and $\bar\nu_e$. The neutrino annihilation {\it number rate} is naturally the same for $\nu_e$ and $\bar\nu_e$.
The use of the average energy in this expression is more debatable than the other approximations made in its derivation. If neutrinos of different energies all have the same angular distribution, $\langle  \omega_{(\nu_i)} \rangle_{\rm pair}$ is the energy-weighted average energy of neutrinos (rather than the number-weighted average energy used in earlier sections).
If neutrinos of different energies have different angular distributions, however, there is no simple way to compute $\langle  \omega_{(\nu_i)} \rangle_{\rm pair}$. Considering that low-energy neutrinos are more likely to come from the accretion disk and thus have a higher annihilation cross-section, it is quite likely that the approximate formula slightly overestimates annihilation rates. Yet, this is probably a small contribution to the error in the computation of $Q^{(+)}_{{\rm pair},\nu_i}$ in a moment scheme.

The main issues with the computation of Eq.~(\ref{eq:Qpair}) in the two-moment formalism are that it relies on M1 estimates of the neutrino pressure tensor, and that it is significantly affected by the over-density of polar neutrinos in the M1 closure. To study these effects, we rewrite Eq.~(\ref{eq:Qpair}) as 
\beq
Q^{(+)}_{{\rm pair},\nu_i} = \frac{C^{\rm pair}_{\nu_i\bar\nu_i} G_F^2}{3\pi \bar h^4 c^3} \langle \omega_{(\nu_i)} \rangle_{\rm pair}  \kappa_{(\nu_i)} E \bar E,
\label{eq:kappa}
\eeq
with $\kappa_{(\nu_i)}$ a dimensionless factor capturing the angular distribution of (anti)neutrinos. As $Q^{(+)}_{{\rm pair},\nu_i} \propto E \bar E$, Fig.~\ref{fig:GlobalDis} provides us with an estimate of the impact on $Q^{(+)}_{{\rm pair},\nu_i}$ of the M1 code's inaccurate values for the neutrino energy density. The M1 code would overestimate annihilation of electron neutrinos by a factor of $\sim 2$, and of heavy-lepton neutrinos by a factor of $\sim 3$.  The geometric factor $\kappa$ has the opposite effect. In the M1 approximation, the assumed distribution function of polar neutrinos is more forward-peaked than what we find with the MC code. As a result, $\kappa$ is significantly underestimated when using the M1 closure. This last effect is shown on Fig.~\ref{fig:PairAnnihilation}: in most of the polar regions, $\kappa$ is larger by factors of $3-5$ in the MC code than in the M1 closure, with peak ratios of $\sim 100$. Accounting for both effects, we estimate that, {\bf given a good estimate of $\langle  \omega_{(\nu_i)} \rangle_{\rm pair}$}, the two-moment code captures the impact of pair annihilations within a factor of $2-3$. 

\begin{figure}
\begin{center}
\includegraphics[width=.99\columnwidth]{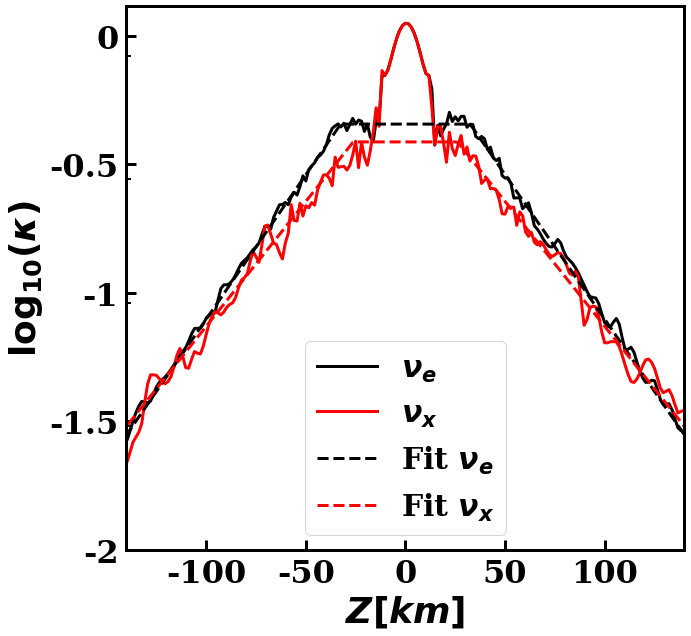}
\caption{Geometric factor $\kappa$ in Eq.~\ref{eq:kappa}, along the polar axis. We show results for electron and heavy-lepton neutrinos, as well as best-fit curves using Eq.~(\ref{eq:kfit}). The fits ignore the high-density regions inside the remnant neutron star, where pair annihilation is a subdominant process.}
\label{fig:KappaFit}
\end{center}
\end{figure}

To help with future computations of the $\nu\bar\nu$ annihilation rate in two-moment simulations, we also provide direct measurements of the geometric factor $\kappa$ in the MC code (Fig.~\ref{fig:PairAnnihilation}). We note that, in the polar regions, $\kappa$ is largely independent of latitude, and mostly depends on the distance to the remnant. In Fig.~\ref{fig:KappaFit}, we show that $\kappa$ is reasonably well fitted by the expression
\beq
\kappa(r) = \min(\kappa_0,A e^{-r/W})
\label{eq:kfit}
\eeq
with $r$ the radius and, for electron neutrinos, $\kappa_0 = 0.53$, $W=37\,{\rm km}$, while for heavy-lepton neutrinos, $\kappa_0=0.45$, $W=43\,{\rm km}$.
In theory, it may be interesting to use these fits in simulations performed with the M1 closure. However, we should note that this will not get rid of errors caused by inaccurate neutrino energy densities in M1 simulations. 

At the very least, we can use our results to estimate the accuracy of existing approximations used to compute the neutrino pair annihilation rate. For example, Fujibayashi et al.~\cite{Fujibayashi:2017puw} use two different methods to compute that rate in their simulations: one using the pressure provided by the M1 closure, and one assuming an isotropic distribution of neutrinos ($\kappa \sim 1.1$).  Our results indicate that the second is a slightly more accurate approximation of $\kappa$ than the first within $50\,{\rm km}$ of the remnant, where most of the pair annihilation energy is deposited. However, even there it is a factor of $\sim 2$ too high, and this error on $\kappa$ acts in the same direction as the error due to the overestimated energy density at the poles. Once both source of errors are taken into account, it appears that consistently using the M1 closure during all calculations is in fact more accurate. 

Finally, we can provide an estimate of the heating rate due to pair annihilation in our simulation. In Fig.~\ref{fig:PairAnnihilation}, we show the heating rate per unit mass. Fujibayashi et al.~\cite{Fujibayashi:2017puw} estimate that the terminal Lorentz factor of the outflows is
\beq
\Gamma_f \sim 1.1 \frac{(Q/\rho)}{10^{24}\,{\rm erg/g/s}} \frac{\tau_{\rm heat}}{1\,{\rm ms}},
\eeq
with $\tau_{\rm heat}$ the time during which the outflows are heated at a constant rate $Q/\rho$. The observed heating rate would make the polar outflows mildly relativistic, as observed in~\cite{Fujibayashi:2017puw}. Pair annihilation is thus important to the dynamics of the outflows. We can also look at the total energy deposition rate in the polar outflows, i.e. in regions with $\theta < 30^o$ and $\rho<10^{10}\,{\rm g/cm^3}$ (the exact value is not very sensitive to changes of the limiting density, even by an order of magnitude). We find an energy deposition $Q_{\rm pair,tot}\sim 3\times 10^{50}\,{\rm erg/s}$. If we assume that the neutrino luminosity decreases on a timescale of $\sim 50\,{\rm ms}$, as in~\cite{Fujibayashi:2017puw}, we find a total energy deposition $\sim 1.5\times 10^{49}\,{\rm erg}$, close to the kinetic energy of the polar ejecta measured in~\cite{Fujibayashi:2017puw} ($\sim 10^{49}\,{\rm erg}$ in their simulation using the M1 closure to compute the annihilation rate). This qualitative agreement is not overly surprising. Using our MC code, we have just argued that an M1 estimate of the annihilation rate should be correct within a factor of $\sim 2-3$.

Overall, we find that multiple approximations made in the M1 scheme create errors of factors of a few in the computation of the $\nu\bar\nu$ annihilation rate. Yet, these errors partially cancel, and we find that a two-moment scheme using the M1 closure and informed by a reasonable estimate of the average neutrino energy can capture the neutrino pair annihilation rate within a factor of $\sim 2-3$. While probably insufficient for detailed studies of the impact of neutrino pair-annihilation on the dynamics of polar outflows, such accuracy is a much more favorable result for the two-moment scheme than one might have assumed before comparison with MC results.

\section{Conclusions}

We perform the first time-dependent, general relativistic, Monte-Carlo radiation transport simulation of neutrinos in the remnant of a binary neutron star merger. While we do not couple the MC evolution to the fluid evolution, we use our results to estimate important sources of errors in more approximate transport algorithms currently used in merger simulations. In particular, we focus on the limitations of the gray two-moment scheme with analytical M1 closure, as implemented in the SpEC code.

We find that the M1 closure is very inaccurate in the low-density polar regions. These regions are of great importance for EM counterparts to neutron star mergers. It is there that the hot, high $Y_e$ material powering optical kilonovae is most likely ejected. Neutrinos in polar regions are also likely to impact the production of short gamma-ray bursts. An important consequence of this inexact closure is that the energy density of neutrinos in the polar regions is strongly overestimated when using the M1 closure, by $\sim 50\%$ for electron-type neutrinos and $100\%$ for heavy-lepton neutrinos. 

The average energy of the neutrinos, on the other hand, is relatively well approximated in the polar regions by two-moment schemes that evolve both the energy and number density of neutrinos (within $\sim 10\%$), but inexact closer to the equatorial plane. We also show that the energy spectrum of electron-type neutrinos is harder than the black-body spectrum usually assumed in the gray M1 scheme.

Combining these various sources of errors, we can estimate that the absorption rates for charged-current reactions responsible for the evolution of the composition of the outflows may be off by factors of $\sim 1.5-2$ in M1 simulations, potentially a fairly significant limitation to our ability to model the composition of the outflows, and thus kilonovae. 

We also consider the impact of the M1 closure on estimates of the $\nu\bar\nu \rightarrow e^+ e^-$ annihilation rate. While two different issues in M1 simulations each induce errors of factors $\gtrsim 2$, these errors partially cancel. An M1 scheme with a good estimate of the average energy of the neutrinos is likely capable of predicting the neutrino annihilation rate within factors of $2-3$. While certainly significant, these errors are smaller than one might have guessed prior to this study. Including pair-annihilation effects within a two-moment scheme probably leads to at least qualitatively correct behavior of the polar outflows.

Another important objective of this simulation is to assess the feasibility of using time-averaged moments computed from a low-resolution MC evolution as closure for the two-moment scheme, thus removing the need to use the approximate M1 closure or to assume a given energy spectrum~\cite{Foucart:2017mbt}.
We find that the time dependence of the neutrino distribution function over the orbital timescale of the remnant is relatively weak, partially justifying the use of moments averaged over timescales comparable to the dynamical timescale of the system. Additionally, our choice to avoid performing MC evolutions in high-optical depth regions (where $\kappa \Delta x \gtrsim 1$) and to instead simply provide boundary conditions approximating a thermal distribution of neutrinos in these regions does not appear to create significant errors, at least when compared with a simulation placing that boundary deeper in the remnant ($\kappa \Delta x \gtrsim 10$). 

We thus estimate that we can provide moments of the neutrino distribution function with statistical noise at the level of a few percents with as little as $\sim 2.5\times 10^7$ MC packets (for a simulation with $2.8 \times 10^7$ finite volume cells!). This indicates that the two-moment scheme with MC closure that we recently proposed~\cite{Foucart:2017mbt} is computationally affordable in simulations of post-merger remnants, and if stable may provide a convenient way to improve upon the standard two-moment algorithm with M1 closure. At the current accuracy of the MC scheme, such a coupled algorithm could in principle reduce numerical errors in the two-moment scheme by an order of magnitude.

\acknowledgments
The authors thank the members of the SXS collaboration for helpful discussions over the course of this project. 
F.F. acknowledges support from NASA through Grant 80NSSC18K0565.
M.D. acknowledges support through NSF Grant PHY-1402916.
H.P. gratefully acknowledges support from the NSERC Canada, the Canada Research Chairs Program and the Canadian Institute for Advanced Research.
L.K. acknowledges support from NSF grant PHY-1606654,
and M.S. from NSF Grants PHY-1708212, PHY-1708213, and PHY-1404569.
L.K. and M.S. also thank the Sherman Fairchild Foundation for their support.
Computations were performed on the supercomputer Briar\'ee from the Universit\'e de Montr\'eal,
managed by Calcul Qu\'ebec and Compute Canada. The operation of these supercomputers is funded
by the Canada Foundation for Innovation (CFI), NanoQu\'ebec, RMGA and the Fonds de recherche du Qu\'ebec - Nature et
Technologie (FRQ-NT). Computations were also performed on the Zwicky and Wheeler clusters at Caltech, supported by the Sherman
Fairchild Foundation and by NSF award PHY-0960291.

\bibliographystyle{iopart-num}
\bibliography{References/References.bib}

\end{document}